# A 3D Strange Attractor with a Distinctive Silhouette.

# The *Butterfly Effect* Revisited

## Safieddine Bouali

University of Tunis, Management Institute
Department of Quantitative Methods & Economics
41, rue de la Liberté 2000 Le Bardo, Tunisia
Safieddine.Bouali@isg.rnu.tn
Phone : +216 79 325 322
Fax : + 216 71 588 487

We propose firstly an autonomous system of three first order differential equations which has two nonlinear terms and generating a new and distinctive strange attractor. Basic attributes of the system are investigated and the full chaotic nature of the system is reported by a positive Lyapunov exponent. Besides, the bifurcation structure of the dynamical system seems analog to the "Ruelle-Takens-Newhouse route to chaos" since the trajectory maps an intricate toroidal domain for the wide range of one parameter. Furthermore, this new 3D chaotic system performs a new feature of the Sensitive Dependency on Initial Conditions (SDIC) popularized as the *Butterfly Effect* discovered by Lorenz (1963). We noticed that the variation of the Initial Conditions for our system leads not only to different attractors but also to a singular phenomenon of overlapped attractors. It constitutes, at best of our knowledge, a novel singularity of the Chaos Theory.

**The *Butterfly Effect* discovered by Lorenz (1963) emphasizing the Sensitive Dependency on Initial Conditions (SDIC):** *Does the flap of a butterfly's wings in Brazil set off a tornado in Texas?* **deserves a revision. We noticed that the variation of the Initial Conditions for the introduced new 3D strange attractor leads not only to different trajectories but also to different attractors with wrapped arrangement. Generally, attractor switch was determined by the Sensitive Dependency of Parameters. When it is derived from SDIC, it constitutes, at best of our knowledge, a novel singularity of the Chaos Theory. In this rapid communication, we propose an autonomous system of three first order differential equations generating a new and distinctive strange attractor with singular features of its SDIC. The general idea behind this research is the detection of an exclusive topology of an attractor with the minimum necessary of nonlinearities by following the *Parsimony law*: "Entities are not to be multiplied beyond necessity."**

**Indeed, at first glance, the new system leads to a typical chaotic attractor exhibiting a pair of double-wings bridged by twice thin paths. Surprisingly, this new 3D chaotic system overlaps as layers, the boundaries of the sub-basins performing wrapped attractors for different initial conditions. In this regard, we can suggest a modification the Butterfly Effect as follows:** *Does the flap of a butterfly's wings in Brazil set off a tornado in Texas?... maybe when it flies over a daisy. But if it flies near the ground, it could set off the freeze of the Great Lakes, as event of a glacial period covering North America.*

## 1. Introduction

In its seminal research, Rössler[1] found a chaotic attractor in the three dimensional phase space with a unique nonlinear term, i.e. a quadratic cross-term, demonstrating the existence of the strangeness of the chaos with a simple nonlinear structure. Comparatively, the Lorenzian attractor[2], the first strange attractor ever discovered, exhibits the deterministic chaos with two quadratic cross-terms.

Monitoring the literature of 3D strange attractors previously published, we notice that chaotic systems built with two quadratic terms, normal and/or cross-product, have been widely studied. Sprott[3] established several classes of chaotic models with these quadratic terms and explored the traits of the discovered attractors. Several other chaotic models integrating two or more quadratic terms are proposed displaying specific strange attractors (see[4-7] and references therein). However, the modelization incorporating only one nonlinear term but in its cubic order constitutes also an interesting approach to detect chaotic attractors. This distinct branch demonstrates to be fruitful to observe rich patterns of strange attractors (for example Bouali[8]).

In this paper, we explore a ramification of this technique since we try to construct intentionally a new 3D chaotic system incorporating a cubic term but coupled to a quadratic cross-term. It is insightful and relevant to build this kind of model to establish a new topology when focusing the minimum necessary of nonlinearities, at best of our knowledge, to ensure distinctive full chaotic patterns.

Obviously, establishing presently a new 3D chaotic behavior with only two nonlinear terms denoted a scientific significance in comparison of a model with a plethora of nonlinearities. This methodological discipline well-known by the expression *Parsimony law* or the *Occam Razor*: "Plurality should not be posited without necessity." In other words: "The principle gives precedence to simplicity; of two competing theories, the simpler explanation of an entity is to be preferred. The principle is also expressed as "Entities are not to be multiplied beyond necessity.""[9].

To this end, the scope of the present research focuses the presentation of another canonical class of chaotic systems, and no having a topologically equivalent structure. Indeed, the proposed new 3D dynamical system overlaps a total of six items in the right-hand side of the three ODEs. It is obtained by the deep change of the model introduced by Bouali[10].

In the section two, the 3D system is described and its basic dynamical behaviors explored. In section three, the attributes of the new silhouette of the four-scroll strange attractor are studied, mainly its transformation when the set of initial conditions are changed. In the final remarks, we point chiefly to its exclusive findings that confirm the singularity of the attractor.

## 2. The 3D Chaotic System

Consider the following three-dimensional system with two nonlinear terms, i.e. a quadratic term and a cubic one, and retaining a simple specification for the third equation:

$$\begin{cases} dx/dt = \alpha x (1 - y) - \beta z \\ dy/dt = - \gamma y (1 - x^2) \\ dz/dt = \mu x \end{cases}$$

where x, y and z are the state variables of the model, and $\alpha$, $\beta$, $\gamma$, and $\mu$, positive parameters. Equilibrium points can be easily found by solving the steady-state condition:

$$dx/dt = dy/dt = dz/dt = 0$$

The origin $S_0$ (x, y, z) = (0, 0, 0) emerges as the unique equilibrium point. Its stability, for the set of parameters $P_0$ ($\alpha$, $\beta$, $\gamma$, $\mu$) = (3, 2.2, 1, 0.001), appears by the linearization of the system through the Jacobian matrix:

$$J = \begin{vmatrix} \alpha (1 - y) & -\alpha x & -\beta \\ 2\gamma xy & -\gamma (1 - x^2) & 0 \\ \mu & 0 & 0 \end{vmatrix} = \begin{vmatrix} 3 & 0 & -2.2 \\ 0 & -1 & 0 \\ 0.001 & 0 & 0 \end{vmatrix}$$

Thus, the characteristic equation $|J - \lambda I| = 0$ gives the polynomial relation for $S_0$ and $P_0$:

$$(1 + \lambda) (\lambda^2 - 3\lambda + 22.10^{-4}) = 0$$

The eigenvalues become:     $\lambda_1 = -1$          $\lambda_2 = 0.001$          $\lambda_3 = 2.999$

The equilibrium $S_0$ is a *saddle focus* since the eigenvalues $\lambda_2$ and $\lambda_3$ are real and positive. The trajectory is driven away from steady-state along with two eigendirections.

The system generates a 3D strange attractor (fig. 1a and b) exhibiting an exclusive silhouette with two juxtaposed double-wings bridged by thin paths. It is easy to remark its invariance under a $\pi/2$ rotation around the Y-axis since it performs an identical shape.

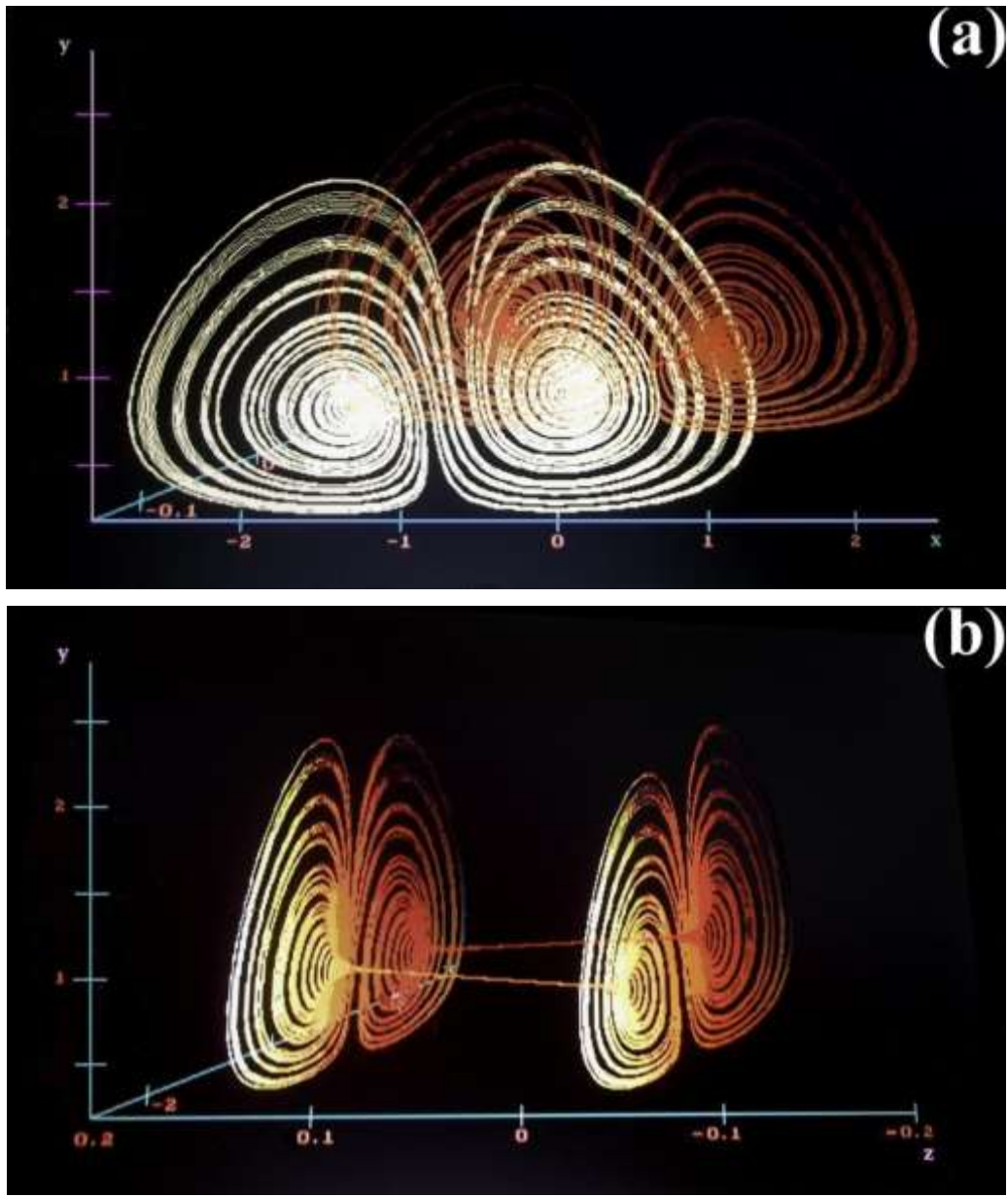

**Fig.1. 3D Phase portrait of the new strange attractor for the parameters**
**$P_0$ (α, β, γ, µ) = (3, 2.2, 1, 0.001) and Initial Conditions $IC_0$ (1, 1, 0).**
**(a)** x− y − z phase space representation. **(b)** z − y − x phase space representation.

The computation of the Lyapunov exponents, with MATLAB using Ode45 routine, and $P_0$ and $IC_0$ performs the following values marking the existence of chaos:

$$L_1 = 0.000733 \qquad L_2 = 0.000530 \qquad L_3 = -0.000592$$

Figure 2(a) displays the evolution of the exponents at the final stage of the numerical calculation, and the figure 2(b) exhibits the peculiar progression of the exponent values.

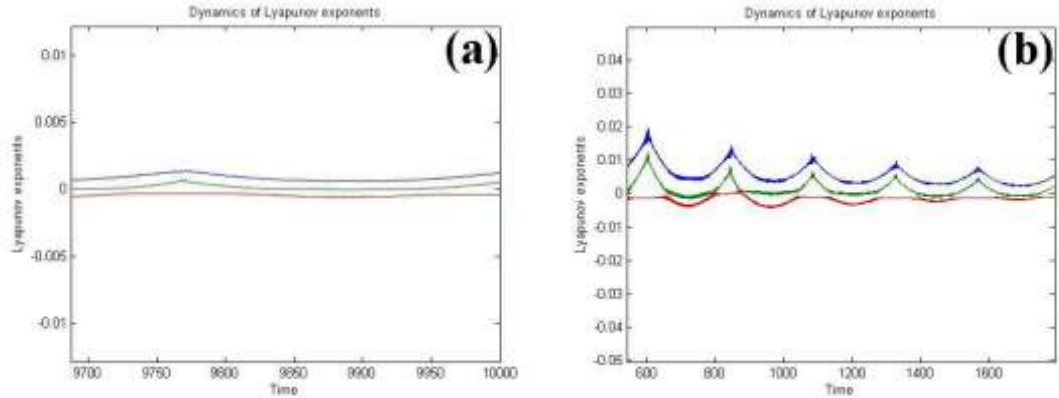

**Fig.2. Lyapunov exponents for the parameters $P_0$ and Initial Conditions $IC_0$.**
(a) Lyapunov exponents $L_1$, $L_2$ and $L_3$ are very close.
(b) The exponent values evolve as waves.

Indeed, peaks depict the chaotic motion of the trajectory inside both the two double-wings.

Furthermore, the measure of the Lyapunov Dimension, i.e. the fractal dimension, by the Kaplan-Yorke[11] method provides $D_{KY} = 2.13$.

On the other hand, the system displays topological mutations when μ constitutes the control parameter. Indeed, for two relatively high values of μ, dynamical trajectory built a domed or an intricate Torus $T^3$ (Fig.3a, and b.) and the existence of such behavior validates the relevance of the "Ruelle-Takens-Newhouse route to chaos"[12].

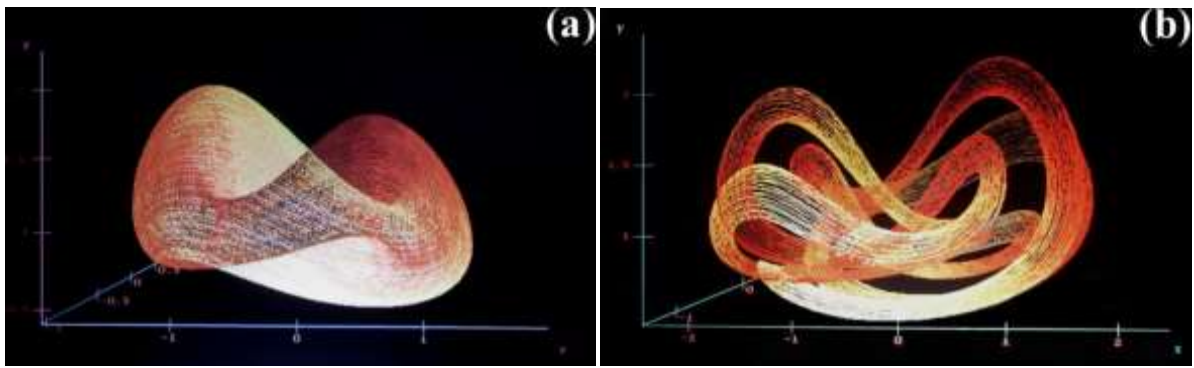

**Fig.3. Torus $T^3$ for Initial Conditions $IC_0$**
(a) with $P_1$ (α, β, $γ_1$, μ) = (3, 2.2, 1, 1.20), and (b) with $P_2$ (α, β, $γ_2$, μ) = (3, 2.2, 1, 1.51)

We noticed that dynamical structures design the precursor of the attractor. Such sequence of transformation constitutes the Sensitive Dependency on Parameters (SDP) as a classical feature of the chaotic systems.

## 3. The Sensitive Dependency on Initial Conditions: The *Butterfly Effect* reconsidered

Surprisingly, the attractor displays also morphological mutations driven not by the SDP but according to the Sensitive Dependency on Initial Conditions. For example for $IC_1$ $(x_1, y_1, z_1)$ = (1, 1, - 0.02), and $IC_2$ $(x_2, y_2, z_2)$ = (1.2, 1, -0.02), the trajectories do not converge to the attractor displayed with $IC_0$ $(x_0, y_0, z_0)$ = (1, 1, 0) in figure 1, as expected in accordance with the nonlinear analysis literature.

These dynamics built different chaotic attractors, and spectacularly, overlapping the strange attractor displayed in the previous section (fig.4). The thin paths connecting the double-wings of the strange attractor are substituted by helical oscillations building tunneled bridges through which the thin paths of the first attractor are threaded.

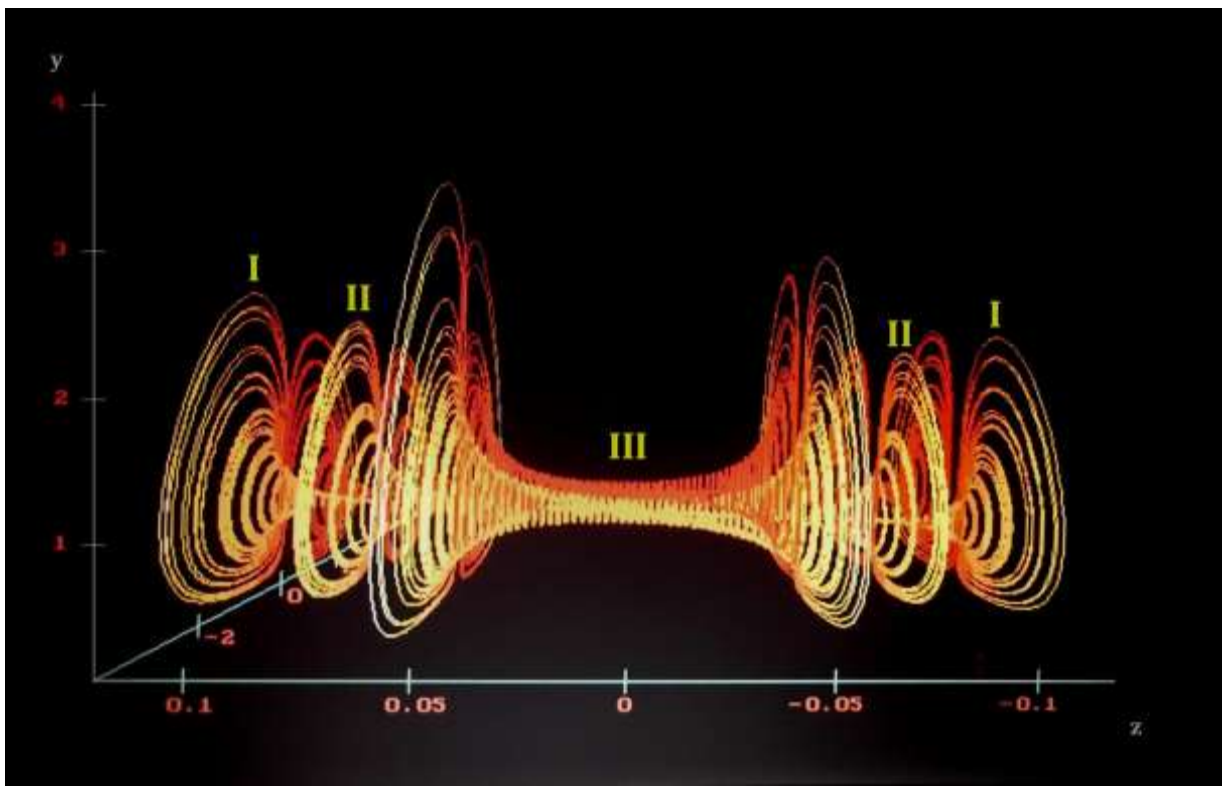

**Fig.4. The embedding phenomenon of the Strange Attractors via the SDIC.**
**Chaotic System with the parameters $P_0$ ($\alpha$, $\beta$, $\gamma$, $\mu$) = (3, 2.2, 1, 0.001).**
**(I)** The strange attractor for $IC_0$ $(x_0, y_0, z_0)$ = (1, 1, 0), **(II)** the strange attractor for $IC_1$ $(x_1, y_1, z_1)$ = (1, 1, - 0.02), and **(III)** the strange attractor for $IC_2$ $(x_2, y_2, z_2)$ = (1.2, 1, -0.02)

Generally, the attractors are simultaneous when the whole basin of attraction is divided in two or several sub-basins separating strictly the attractors, i.e. every attractor appears in its specific basin. In the present case, the embedding phenomenon of the attractors seems violate this principle. However, the numerical exploration leads to the detection of a particular folding of the sub-basin boundaries that are arranged as wrapped layers. Each layer gathers a stratum of initial conditions confining its specific strange attractor. The whole basin of attraction assembles a series of wrapped sub-basins locking in each one its strange attractor.

It seems that phase space attribute derives from the divergence nature of the whole vector field of the chaotic system. Indeed, we discuss its formulation:

**div V** = Tr(J) = α (1 − y) − β (1 − $x^2$)

for α and β positive values. Dissipativity and volume contraction of the flow is accurately identified when:

**div V** < 0 then -α (1 + y) < -β (1 + $x^2$)

The restriction emerges as follows:

- (α (1+ y) − β) /β] $^{0.5}$< x < [(α (1+ y) − β) /β] $^{0.5}$, and assuming the condition:

y > (β − α)/ α

In this precise case, orbits near the chaotic attractor are ultimately limited to a specific fractal-dimensional subspace of zero volume. Otherwise, it appears that the x and y initial conditions breaking the previous constraints found a volume conservation or expansion leading to the phenomenon of overlapped attractors.

At best of our knowledge, this SDIC consequence composes a novel singularity of the Chaos Theory. Consequently, the expression popularizing the Lorenzian *Butterfly Effect*: *"Does the flap of a butterfly's wings in Brazil set off a tornado in Texas?"* deserves a revision.

Indeed, in this meteorological instance, the attractor of the earth climate involves a broad variety of status, between them: cloudy, sunny days, storms, rain, etc., and in line with the SDIC concept, the butterfly flaps initiate one arrangement of these statuses amongst an infinite number. Weather becomes unpredictable.

Our results broke this inspiring vision.

A single leap of the butterfly to a different location allows to its flaps the ability to drive the meteorological system of the Earth from the mentioned attractor governing the earth climate to another involving, for example, a series of icy burst, blast of freezing wind, glacial gale, heavy snowfall, etc. Obviously, such variety of status composes the weather of a glacial period and its infinite number of arrangements could be reported in a theoretical chaotic attractor. The stylized facts of the Butterfly Effect are now expanded not only to the long term unpredictability of the weather for the current meteorological climate but to all climate statuses , from the proterozoic period to the present regime. Climate becomes unpredictable. Indeed, the appendix proposed to the Butterfly Effect can be formulated as follow*: Does the flap of a butterfly's wings in Brazil set off a tornado in Texas?... maybe when it flies over a daisy. But if it flies near the ground, it could set off the freeze of the Great Lakes, as event of a glacial period covering North America.*

## 4. Final Remarks

The introduced chaotic system of three autonomous nonlinear equations embeds two nonlinearities, namely xy, the quadratic term and, $yx^2$ the cubic cross-product term amongst a total of six items in all equations. This restricted number of nonlinearities constitutes currently an eminent trait of the model since the exploration of low dimensional chaos has been broadly carried out. Besides, the model contains a unique equilibrium, unstable, revealing an Index-2 of stability and displaying a very rich repertoire of traits, predominantly its distinctive and exclusive new silhouette. Moreover, Dominant Lyapunov exponent is positive signifying the presence of chaotic patterns.

Our rapid presentation pointed chiefly to an essential feature of the model. An attribute of extreme Sensitive Dependency on Initial conditions that are shown through concomitant strange attractors is revealed. For the same specification of parameters, the phase space envelops several wrapped sub-basins of attraction. Definitely, Butterfly Effect is not restricted to a sole climate but concerns all the weather statuses of the long-ago known climates.

Obviously, this phenomenon requires further analytical descriptions to complete its mathematical treatment.